**Non-topological Origin of the Planar Hall Effect in Type-II Dirac Semimetal NiTe$_2$**

Qianqian Liu, Bo Chen, Boyuan Wei, Shuai Zhang, Minhao Zhang, Faji Xie, Muhammad Naveed, Fucong Fei*, Baigen Wang[†], Fengqi Song [‡]

*National Laboratory of Solid State Microstructures, Collaborative Innovation Center of Advanced Microstructures, and College of Physics, Nanjing University, Nanjing 210093, China*

# ABSTRACT

Dirac and Weyl semimetals are new discovered topological nontrivial materials with the linear band dispersions around the Dirac/Weyl points. When applying non-orthogonal electric current and magnetic field, an exotic phenomenon called chiral anomaly arises and negative longitudinal resistance can be detected. Recently, a new phenomenon named planer Hall effect (PHE) is considered to be another indication of chiral anomaly which has been observed in many topological semimetals. However, it still remains a question that is the PHE only attributed to chiral anomaly? Here we demonstrate the PHE in a new-discovered type-II Dirac semimetal $NiTe_2$ by low temperature transport. However, after detailed analysis, we conclude that the PHE results from the trivial orbital magnetoresistance. This work reveals that PHE is not a sufficient condition of chiral anomaly and one need to take special care of other non-topological contribution in such studies.

# I. INTRODUCTION

Topological semimetals (TSMs) are currently creating a surge of research activities in condensed matter physics due to their novel properties. Multiple series of TSMs have been theoretically predicted as well as experimentally discovered and researched, such as Dirac semimetals, Weyl semimetals, nodal-line semimetals, etc. Most of the discovered TSM materials, such as $Cd_3As_2$, $Na_3Bi$, TaAs [1-10], hold linear cone-shaped band dispersions. Point-like Fermi surface will be formed when the Fermi level aligns with the Dirac/Weyl points. Meanwhile, in some special TSM systems, such as $WTe_2$, LaAlGe, $VAl_3$ and $PtSe_2$ family [11-19], the Lorentz invariance is broken and the Dirac/Weyl cones are strongly tilted along a certain momentum direction. A pair of electron and hole Fermi pockets touching each other at Dirac/Weyl point is formed because of the tilting. This unique Fermi surface configuration is distinct from the traditional cases and these materials with Lorentz invariance breaking are called type-II TSMs. Accordingly, the previous mentioned TSMs with point like Fermi surface are named type-I TSMs. Special Fermi surface configurations in type-II TSMs are expected to result in many special phenomena such as Klein tunneling in momentum space [20], anisotropic electrical transport [21], angle-dependent chiral anomaly [22], etc. Among these peculiar characteristics, chiral anomaly has long been concerned and widely investigated in many TSMs. In Weyl semimetals, when applying external magnetic field and electrical current which are parallel to each other, electrons on the lowest Landau level will be pumped from one Weyl cone to another with opposite chirality, resulting in a chiral current. Thus negative magnetoresistance (NMR) can be experimentally observed when increasing the magnetic field [23,24]. Same scenario appears in Dirac semimetals as a Dirac point splits into two Weyl points with opposite chirality when applying magnetic field [24]. However, the NMR phenomenon can't be used as a solid evidence to judge the existence of chiral anomaly as other unexpected effects such as extrinsic current jetting effects may cause similar phenomenon [25,26]. Recent theories suggest that the planar Hall effect (PHE) in topological semimetals is another transport evidence of chiral anomaly which magnetic field rotates within the *x-y* plane when measuring electrical transport by a standard Hall-bar configuration (diagrammatically displayed in Fig. 3(b)) [27,28]. So far, PHE has been observed in both type-I and type-II TSMs such as $ZrTe_5$, $Cd_3As_2$, GdPtBi, $VAl_3$, $MoTe_2$, and so on [29-33]. It seems a general phenomenon which can be detected in many different kinds of TSMs. However, the origin of PHE phenomenon in these materials is still not clear and cannot curtly pin down to chiral anomaly. Several other effects such as magnetic order, spin-orbital coupling and in-plane orbital magnetoresistance may also cause similar phenomenon which need to be carefully distinguished.

Here we perform an observation of the PHE and anisotropic longitudinal magnetoresistance (AMR) in $NiTe_2$ which is a new discovered type-II Dirac semimetal. The Dirac points of $NiTe_2$ are close to the Fermi surface [34] and PHE induced by chiral anomaly is expected to be detected in this material. After the systematic study of the angle-dependence of planar Hall resistance and longitudinal magnetoresistance (MR) under different magnitude of magnetic field and various temperatures, however, we conclude that the PHE in NiTe$_2$ originates from the trivial in-plane orbital magnetoresistance rather than expected chiral anomaly. Our result takes an example that PHE measured in topological materials may not be caused by chiral anomaly. When PHE is detected in topological semimetals, one need to be more cautious to attribute it to chiral anomaly or nontrivial Berry phase, and further detailed investigations are essential.

## II. RESULTS AND DISCUSSION

### A. Crystal growth and characterization

Single crystals of NiTe$_2$ are grown by self-flux method. Nickel powder (from Aladdin) and tellurium shot (from Alfa Aesar) were mixed with a molar ratio of 1:15 in a glove box and sealed in an evacuated quartz tube. The quartz tube was then heated to 700 ℃ quickly in a muffle furnace and kept at this temperature for 12 h, before being slowly cooled down to 500 ℃ (3 ℃/h), The excess amount of Te was centrifuged at 500 ℃. Millimeters size of crystals with shiny surface can be obtained. NiTe$_2$ crystal holds a CdI$_2$-type trigonal structure with $P\bar{3}m1$ space group. It is a layered material and the single layers of NiTe$_2$ stack along the *c*-axis, as shown in Figure. 1(a). Fig. 1(b) displays the energy dispersive spectrum (EDS) of one typical crystal. One can clearly see the Ni and Te peaks with a perfect atomic ratio of 1:2. Clear (00n) diffraction peaks of the NiTe$_2$ can be seen in the single crystal X-ray diffraction (XRD) pattern in Fig. 1(c), and there are no other impurity peaks, indicating the high quality of the crystals. In Fig. 1(d), the resistance versus temperature exhibits a metal property, and the residual resistance ratio (RRR) is 111 at zero field. Fig. 2(a) depicts the magnetoresistance curves under various temperatures when the magnetic field perpendicular to the sample surface. As noticed, the MR ratio ($(\rho(B)-\rho(0))/\rho(0)$) of the sample is relatively larger, reaching as high as 400% at 2 K, 10 T. Besides, the MR at low temperature is quite linear, which is consistent with previous report [34]. In Fig. 2(b), the MR ratio under different temperatures is displayed, which is reduced by 14 times from 2 K to 50 K when $B$ = 10 T. Fig. 2(c) represents the longitudinal resistance varies with different angles when magnetic field rotating out of the *x-y* plane as displayed by the inset cartoon, and the out-of-plane anisotropic MR can be obviously seen. Fig. 2(d) shows the Hall resistivity of the sample. The bend Hall curves indicate the multi-band electrical transport channels in NiTe$_2$, which is consistent with the band structure calculations of this material [34].

### B. Measurement of planer Hall effect

In the next step, we rotate the direction of magnetic field into the *x-y* plane (*a-b* plane of the NiTe$_2$ crystal) and measure the in-plane PHE and AMR properties. Quantitatively, the resistivity tensor considering chiral anomaly can be written as follows [27]:

$$\rho_{yx}^{PHE} = -\Delta\rho^{chiral} \sin\theta \cos\theta \qquad (1)$$

$$\rho_{xx} = \rho_{\perp} - \Delta\rho^{chiral} \cos^2\theta \qquad (2)$$

where $\rho_{yx}^{PHE}$ represents the in-plane hall resistivity which directly shows up the PHE. The $\Delta\rho^{chiral} = \rho_{\perp} - \rho_{\parallel}$ is the chiral anomaly induced resistivity anisotropy, $\rho_{\perp}$ and

$\rho_\parallel$ represent the resistivity with the magnetic field perpendicular (90°) and parallel (0°) to the electric current respectively. $\rho_{xx}$ is the AMR when magnetic field rotating in the sample plane as displayed in the inset of Fig. 3(b). Both $\rho_{yx}^{PHE}$ and $\rho_{xx}$ hold a period of 180°. In particular, for $\rho_{yx}^{PHE}$, the valleys appear at 45° and 235° whereas the peaks appear at 135° and 315°, which is distinct from the angular dependence measured in an ordinary Hall effect when magnetic field perpendicular to the sample *x-y* plane, which holds a period of 360°. The device configuration is displayed by the schematic diagram in Fig. 3(b), with two longitudinal electrodes measuring the in-plane longitudinal resistance and other two lateral electrodes measuring the PHE signal. The applied magnetic field rotates within the sample surface and $\theta$ represents the angle between the in-plane magnetic field and the applied electric current. It is worth mentioning that in practical cases, there are several misalignments which may influence the PHE measurement. Firstly, the magnetic field commonly does not rotate in the plane strictly and a residual out-of-plane component may exist, which will generate a normal Hall signal mixing into the real in-plane response and affect the measurement of PHE. To rule out this contribution, we measured the transport signal under both positive and negative field and average the positive/negative data, as the normal Hall signal is opposite when reversing the field direction while the PHE signal holds the same. Secondly, if two Hall electrodes are deviated each other longitudinally, in-plane longitudinal AMR and out-of-plane longitudinal MR signals will be induced. The former one possesses a $\cos^2\theta$ angle relation which can be ruled out if processing the data by formula: $\rho_{yx} = (\rho_{yx}(\theta) - \rho_{yx}(\pi - \theta))/2$. The latter out-of-plane longitudinal MR contribution is troublesome as it is $\cos^2(\theta+\delta)$ dependence and the phase shifting $\delta$ is stochastic depending on the geometrical relationship between the current and magnetic field direction. We measured several samples and find that the PHE data always show the peak near 135° (315°) and the valley near 45° (225°). Therefore we believe that this out-of-plane MR with random phase is small and can be ignored. After considering all these items, we plot the intrinsic PHE signal in Fig. 3(a) and one can see that experimental data demonstrate a period of 180° period, with valleys at 45° and peaks at 135°, which fit well with the fitting curves (red lines) based on equation (1). We plot the amplitude of the planar Hall coefficient $\Delta\rho_{yx}$ versus magnetic field in Fig. 3(b). The solid line is the power law fitting curve of the data points, showing the field dependence with an exponent 1.41. According to the previous theoretical calculation, if the PHE signal purely comes from chiral anomaly, $\Delta\rho_{yx}$ should be proportional to $B^2$ [28]. The deviation of the exponent indicates that the PHE may be not originated from pure chiral anomaly. To demonstrate the behavior of PHE in more detail, we research the temperature-dependent amplitude of the PHE at 14 T. The amplitude persist until 150 K as shown in Fig. 3(c) and the planar Hall coefficients $\Delta\rho_{yx}$ at different temperatures are extracted from fitting data in Fig. 3(d) by equation (1). It is clear that $\Delta\rho_{yx}$ decreases quickly from 2 K to 60 K, and the rate of decline gradually slows down at higher temperature. It is reduced by almost 5 times at 150 K comparing the PHE signal at 2 K.

### C. Origin of measured planer Hall effect

From the PHE measurement and the analysis of it mentioned above, though it fit well with equation (1), it is still hard to figure out whether the measured PHE signal truly comes from chiral anomaly or it is just a trivial phenomenon caused by some other factors. Therefore, we next focus on the in-plane longitudinal AMR signals measured simultaneously. Fig. 4(a) displays the measured the in-plane AMR under different fields at 2 K. As the field rotates, a pronounced modulation of longitudinal resistivity is observed, with a period of 180º. Using equation (2) to fit the experimental data, one can find that the fitting curves (red lines) have a good agreement with the experiment. The $\Delta\rho_{xx}$ at different fields which can be further fitted to the power law curve with $\Delta\rho_{xx} \propto B^{1.41}$, as shown in Fig. 4(b), the fitting exponent is consistent with the result in Fig. 3(b), confirming the reliability and analyses of our measurements. The PHE and AMR curves show the $\sin\theta\cos\theta$ and $\cos^2\theta$ dependence respectively and both consistent with the equation described by the pure chiral anomaly, however, one can find that in Fig. 4(a), when magnetic field parallel to the electrical current (0° and 180°), the longitudinal resistivity increases when increasing the magnetic field and exhibits a positive magnetoresistance. This result is in stark contrast to the negative magnetoresistance caused considering chiral anomaly. As mentioned above, equation (1) and (2) are ideal equations which only take chiral anomaly into consideration. Namely, for $\rho_{xx}$, negative magnetoresistance should be displayed. In addition, in the pure chiral anomaly case, the resistivity when magnetic field perpendicular to current ($\rho_\perp$) should keep constant and the resistivity when magnetic field parallel to current ($\rho_\parallel$) should decrease as $B$ field increases [29,32]. In Fig. 4(c), we plot the extracted $\rho_\perp$ and $\rho_\parallel$ from Fig. 4(a) with the vary of $B$ field. It can be seen that $\rho_\perp$ changes linearly with $B$ and $\rho_\parallel$ changes relatively slowly which satisfies the relation of the polynomial. Both $\rho_\perp$ and $\rho_\parallel$ increase when increasing the magnetic field, which is distinct from the case of chiral anomaly.

As a result, we consider that the chiral anomaly is not the origin of our measured PHE results.

To unveil the real factors leading to this kind of PHE signal, we temporarily set aside the specific system with chiral anomaly and take a general electric conductor system into consideration. If the in-plane magnetic field causes the in-plane anisotropic resistivity in this system and the resistivity tensor can be written as:

$$\begin{pmatrix} E_{x'} \\ E_{y'} \end{pmatrix} = \begin{pmatrix} \rho_\parallel & 0 \\ 0 & \rho_\perp \end{pmatrix} \begin{pmatrix} j_{x'} \\ j_{y'} \end{pmatrix} \qquad (3)$$

where the $x'$ direction is the direction of $B$ field to be applied and the $y'$ direction is perpendicular to the field. If we take the sample itself as the coordinate system, after executing a standard coordinate transformation procedure, formula (3) becomes:

$$\begin{pmatrix} E_x \\ E_y \end{pmatrix} = \begin{pmatrix} \cos\theta & -\sin\theta \\ \sin\theta & \cos\theta \end{pmatrix} \begin{pmatrix} \rho_\parallel & 0 \\ 0 & \rho_\perp \end{pmatrix} \begin{pmatrix} \cos\theta & \sin\theta \\ -\sin\theta & \cos\theta \end{pmatrix} \begin{pmatrix} j_x \\ j_y \end{pmatrix}$$
$$= \begin{pmatrix} \rho_\parallel \cos^2\theta + \rho_\perp \sin^2\theta & (\rho_\parallel - \rho_\perp)\sin\theta\cos\theta \\ (\rho_\parallel - \rho_\perp)\sin\theta\cos\theta & \rho_\parallel \sin^2\theta + \rho_\perp \cos^2\theta \end{pmatrix} \begin{pmatrix} j_x \\ j_y \end{pmatrix} \qquad (4)$$

If we determine the $x$ direction is parallel to the electric current, i.e. $j_y = 0$, and $\theta$ is the angle between applied current and $B$ field. Then we can get:

$$\rho_{yx} = E_y / j_x = -(\rho_\perp - \rho_\parallel)\sin\theta\cos\theta \qquad (5)$$

$$\rho_{xx} = E_x / j_x = \rho_\perp - (\rho_\perp - \rho_\parallel)\cos^2\theta \qquad (6)$$

One may find that the form of equation (5) and (6) is exactly the same as equation (1) and (2). That is to say, as long as the resistivity is anisotropic when applying in-plane magnetic field, the PHE can be observed. The anisotropic resistance may be caused by various reasons, not only chiral anomaly, but also classical orbital magnetoresistance, strong spin-orbital coupling in magnetic system, etc. The similar derivation process is also be demonstrated in previous report [35], and the PHE has been experimentally reported not only in TSMs but also in topological insulators [36,37], magnetic material (Ga,Mn)As Devices [38], trivial metal Bismuth [35] and so on. NiTe$_2$ is a non-magnetic bulk 3D material with complicated Fermi surface. The anisotropic orbital magnetoresistance is pretty common in this kind of system. In addition, the out-of-plane anisotropic magnetoresistance is also clear demonstrated by our experiment (Fig. 2(c)). There are reasons to attribute PHE in NiTe$_2$ to its in-plane anisotropic orbital magnetoresistance. For further proofing our opinion, referring to a recent report giving a plausible criterion for chiral anomaly [35], we plot the amplitude of PHE vs in-plane AMR of NiTe$_2$ with $\theta$ as a parameter under a specific magnetic field in Fig. 4(d). The parametric plot pattern of the system in which the PHE is dominated by chiral anomaly, such as Na$_3$Bi and GdPtBi, are concentric around the center [35]. In contrast, in our measurement of NiTe$_2$, as $B$ field increases, the orbits evolve to form a "shock-wave" pattern, which is a typical exemplification with the absence of chiral anomaly. Our point of view that the PHE of NiTe$_2$ is not from chiral anomaly or nontrivial Berry curvature is thus confirmed.

### III. CONCLUSION

We measure the PHE in the type-II Dirac semimetal NiTe$_2$. By carefully analyzing the data received, we find out that the PHE in NiTe$_2$ does not originate from expected chiral anomaly. We find that as long as the resistivity of a solid system is anisotropic when applying an in-plane magnetic field, PHE signal can be detected. We believe the PHE we measured in NiTe$_2$ stems from its in-plane orbital resistance. Our result takes an example that PHE measured in topological materials caused by a non-topological origin. In addition, it is necessary to point out that PHE measured in topological materials when negative longitudinal resistance is absent cannot be originated from chiral anomaly or nontrivial Berry phase. In such studies, special care is needed and one should take both PHE and negative longitudinal resistance into consideration to distinguish the real origin of the PHE signals.

**ACKNOWLEDGEMENTS**

The authors gratefully acknowledge the financial support of the National Key R&D Program of China (2017YFA0303203), the National Natural Science Foundation of China (91421109, 91622115, 11522432, 11574217, U1732273 and U1732159), the Natural Science Foundation of Jiangsu Province (BK20160659), the Fundamental Research Funds for the Central Universities, and the opening Project of the Wuhan National High Magnetic Field Center.

**FOOTNOTES AND REFERENCE CITATION**

\* Corresponding authors.

feifucong@nju.edu.cn

† Corresponding authors.

bgwang@nju.edu.cn

‡ Corresponding authors.

songfengqi@nju.edu.cn

**FIGURE CAPTIONS**

FIG. 1. The crystal growth and characterization of the NiTe$_2$ single crystals. (a) The CdI$_2$-type crystal structure of NiTe$_2$. (b) The EDS spectrum of synthesized NiTe$_2$ crystal, which demonstrates a stoichiometric ratio. (c) The single crystal X-ray-diffraction data of the (00n) surfaces of the sample. The inset is the optical micrograph of the NiTe$_2$ crystal, the scale-bar is 4 mm. (d) The resistivity varies with temperature at zero field.

FIG. 2. The transport properties of the NiTe$_2$ single crystals. (a) The MR ratio at different temperatures when the magnetic field perpendicular to the *x-y* plane. (b) The MR ratio at different temperatures. (c) The longitudinal resistance varies with different angles when magnetic field rotating out of the *x-y* plane ($T = 2$ K). The inset displays the device configuration for out-of-plane transport measurement. (d) The Hall resistivity measured at different angles when magnetic field rotating out of the *x-y* plane ($T = 2$ K).

FIG. 3. Planar Hall effect measurement in NiTe$_2$. (a) The and fitting curves. ($T = 2$ K. (b) The $\Delta\rho_{yx}$ from the equation (1) varies with magnetic field ($T = 2$ K). The inset shows the schematic of the measurement configuration. (c) The planar Hall resistance changes versus angle $\theta$ at different temperatures ($B = 14$ T). (d) The extracted parameter $\Delta\rho_{yx}$ varies with temperature ($B = 14$ T).

FIG. 4. Non-topological origin of the PHE in NiTe$_2$. (a) Measured in-plane AMR versus angle $\theta$ at various fields ($T = 2$ K). Solid red curves represent the fittings to using equation (2). (b) The amplitude of AMR varies with magnetic field at $T = 2$ K. The solid curve is the power law fit curve for the experimental data points. (c) $\rho_\perp$ and $\rho_\parallel$ extracted from the experimental data in panel (a). The orange and blue solid curve represent the power law fit curves for $\rho_\perp$ and $\rho_\parallel$, respectively. (d) The orbits obtained by plotting $\rho_{xx}$ and $\rho_{yx}$ with angle $\theta$ as the parameter at specific magnetic fields. The orbits evolve to form a "shock-wave" pattern, indicating the absence of chiral anomaly.

**FIGURES**

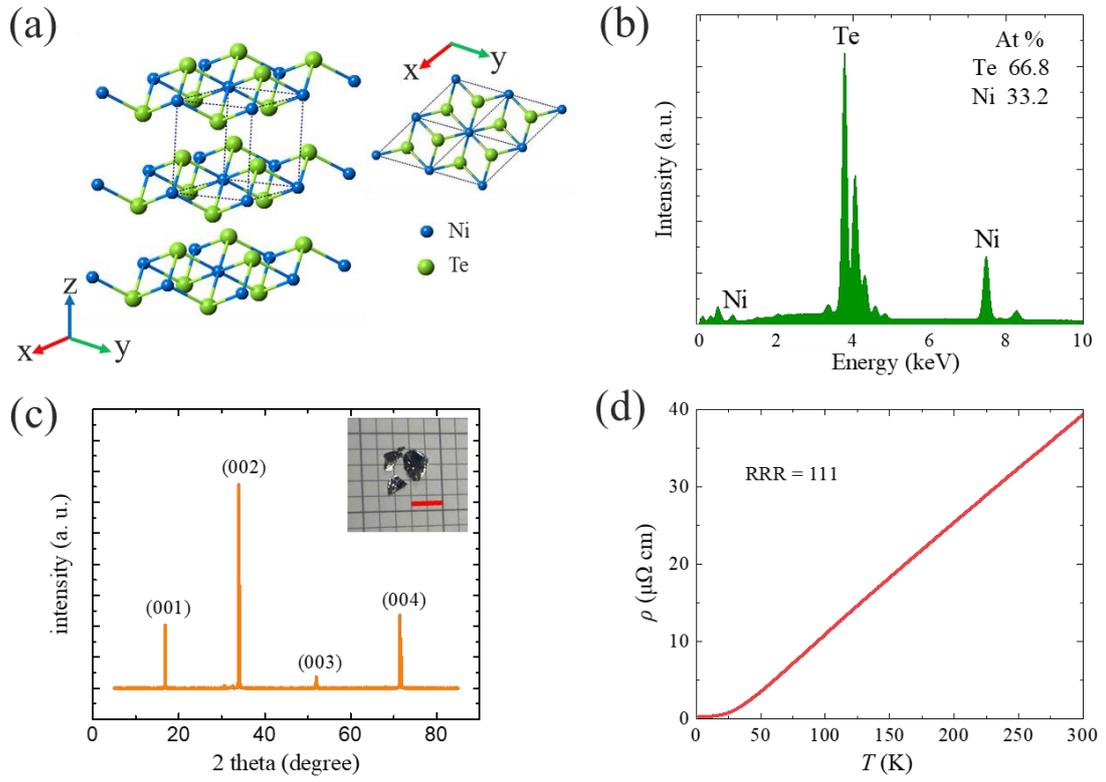

FIG. 1. The crystal growth and characterization of the NiTe$_2$ single crystals.

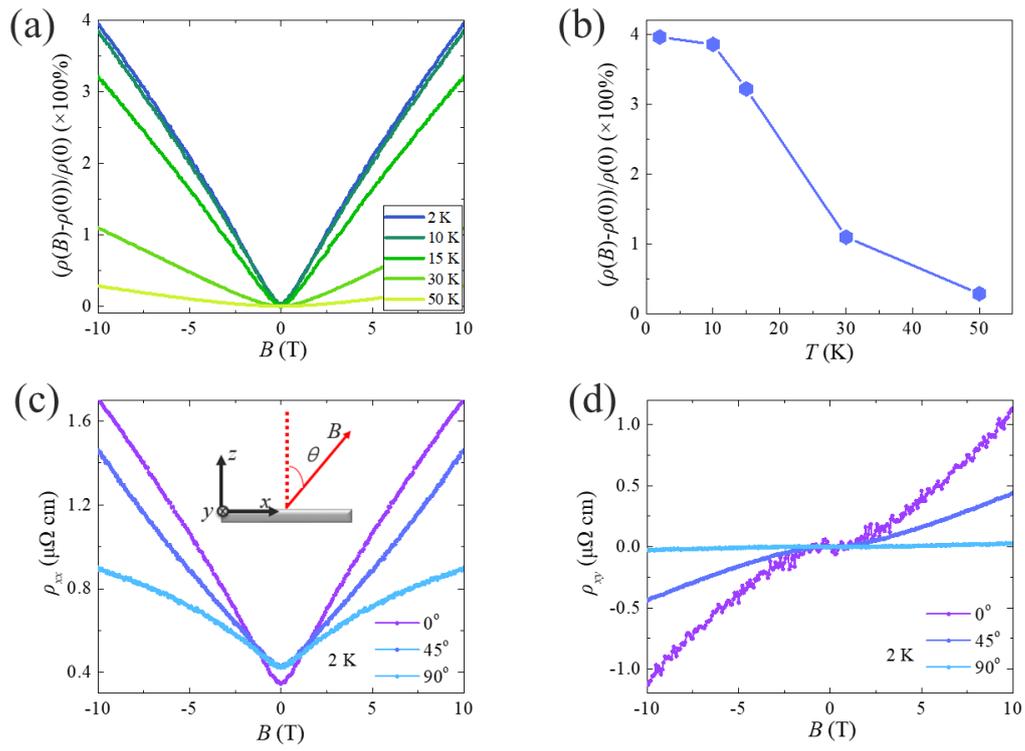

FIG. 2. The transport properties of the NiTe$_2$ single crystals.

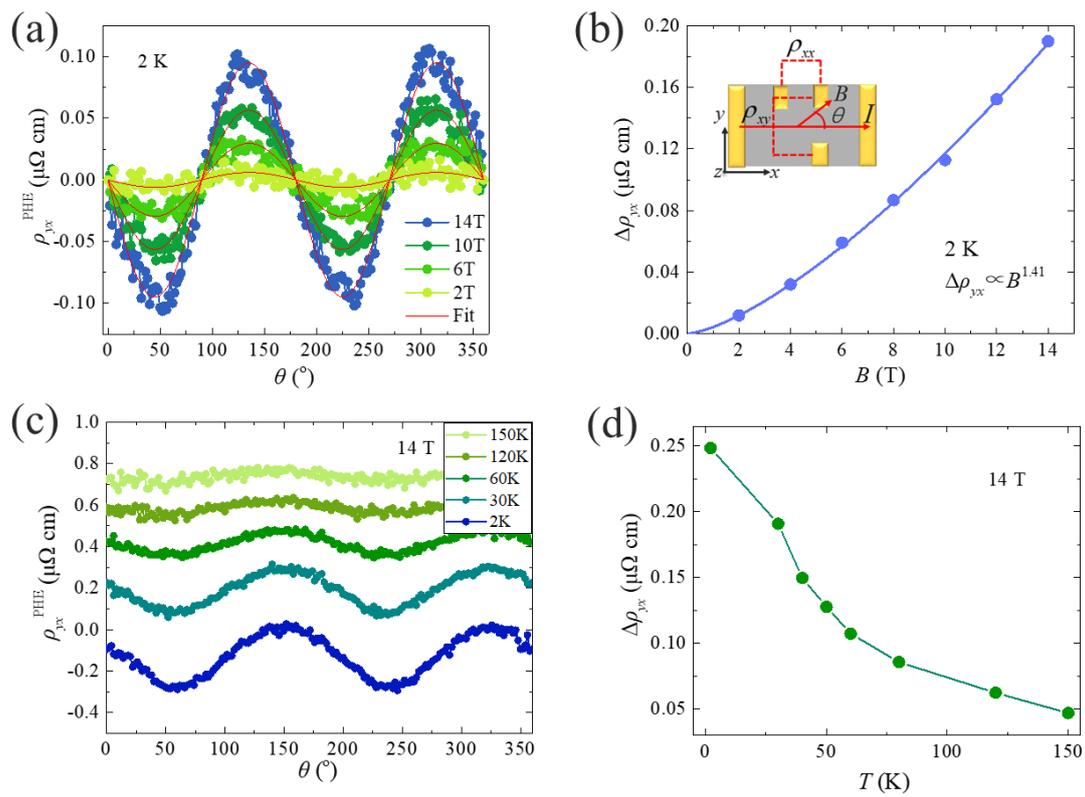

FIG. 3. Planar Hall effect measurement in NiTe$_2$.

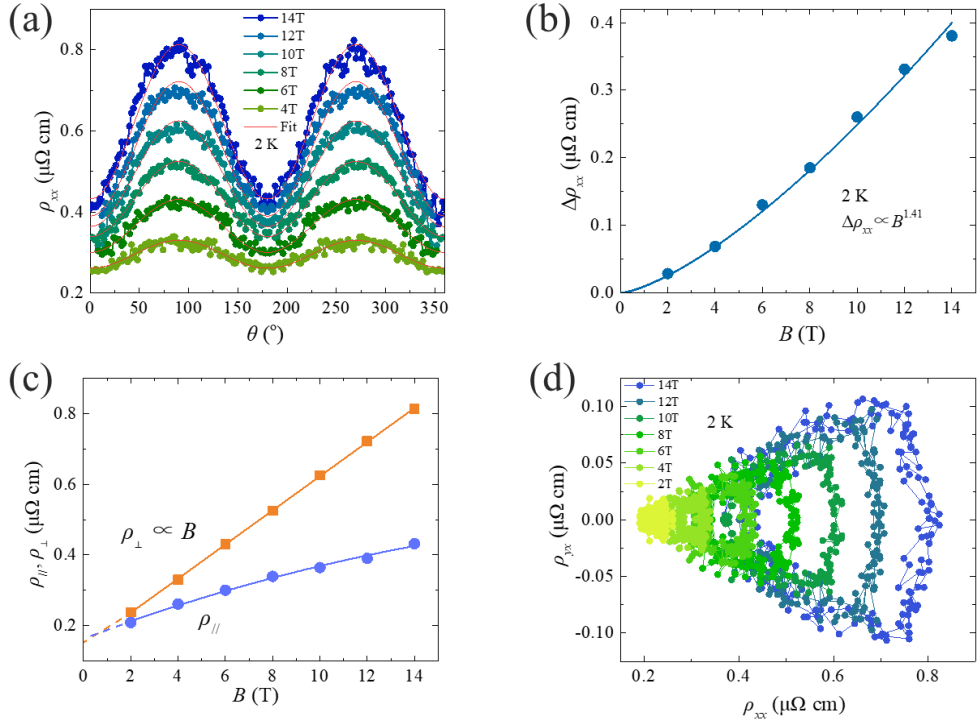

FIG. 4. Non-topological origin of the PHE in NiTe$_2$.